\documentclass[12pt,letterpaper]{article}

\usepackage[T1]{fontenc}
\usepackage[utf8]{inputenc}
\usepackage[english]{babel}
\usepackage[compact,raggedright,small]{titlesec}
\usepackage[affil-it]{authblk}
\usepackage[runin]{abstract}
\usepackage{blindtext}
\usepackage{bm}

\usepackage{graphicx}
\usepackage{color}
\usepackage{epstopdf}

\usepackage[bitstream-charter]{mathdesign}
\usepackage[T1]{fontenc}

\usepackage[numbers]{natbib}
 
\makeatletter

\newcommand{\z}{Z\kern-0.45emZ}
\newcommand{\R}{I\kern-0.3emR}
\newcommand{\vi}{I\kern-0.3emB}
\newcommand{\1}{I\kern-0.3emI}
\newcommand{\e}{I\kern-0.3emE}
\newcommand{\E}{I\kern-0.3emE}
\newcommand{\N}{I\kern-.35emN}
\newcommand{\ind}{I\kern-0.3emI}
\newcommand{\p}{I\kern-0.3emP}
\renewcommand{\P}{I\kern-0.3emP}

\newcommand{\be}{\begin{equation}}
\newcommand{\ee}{\end{equation}}
\newcommand{\bea}{\begin{eqnarray}}
\newcommand{\eea}{\end{eqnarray}}

\addto{\Affilfont}{\small}

\date{\small (Dated February 10, 2015)}
\title{Backward Renormalization Priors and the
Cortical Source Localization Problem with EEG or MEG }

\author[1,2]{Leonardo S Barbosa}
\author[1]{Nestor Caticha}

\affil[1]{Instituto de Física, Universidade de S\~ao Paulo}
\affil[2]{Laboratoire de Science Cognitive et Psycholinguistique, Ecole Normale Supérieure}
\begin{document}
  \maketitle

  \begin{abstract}
    We study source localization from high dimensional M/EEG data by extending
    a multiscale method based on Entropic inference
    devised to increase the spatial resolution of inverse problems .
    This method is used to construct informative  
    prior distributions in  a manner inspired 
    in the  context of fMRI (Amaral et al 2004 \cite{Amaral2004}) .
    We construct a  set of renormalized lattices 
    that approximate the cortex region where the source activity is located and
    address the related problem
    of defining the relevant variables in a coarser scale representation of the cortex. The priors 
    can be used in conjunction with other Bayesian methods such as the 
    Variational Bayes  method (VB, Sato et al 2004 \cite{Sato2004}).
    The central point of the algorithm 
    is that it uses  a posterior obtained at a coarse scale 
    to induce a prior at the next
    finer scale stage of the problem. 
    We present results which suggest,
    on simulated data, that this way of including prior information is a useful aid for the source location problem. This is judged by the rate and magnitude 
    of errors in source localization. 
    Better convergence times are also achieved. We also present
    results on public data collected during a face recognition task. 
  \end{abstract}


\pagebreak

\section{Introduction}
Designing improved methods of inference by  updating probabilities
as new information is aquired, depends partly  on the development of 
ideas of how to incorporate prior information.  
In this paper we concentrate on the construction of  prior distributions
for the localization of cortical dipole sources in EEG or MEG 
high resolution experiments. 
Several groups have presented very encouraging results with 
respect to source localization (for a review see \cite{Ramirez2008}). 
After data aquisition, the source localization analysis
 problem is divided into three essentially separate problems: 
first, construction of the prior; second, of the likelihood or model and
third,
of the algorithm to extract the information from the resulting posterior.
Several such information 
extraction algorithms have been systematically analyzed by Wipf
and  Nagarajan
\cite{Wipf2009} in a useful unification that simplifies the literature, since
different Bayesian approaches differ only in the
available information.
Sato {\it et al} \cite{Sato2004} have tackled the 
construction of empirical priors. 
Our approach, which can also focuses on an empirical prior is based on 
a previously introduced multiscale-prior method \cite{Caticha2015} which formalized ideas 
applied in the context of fMRI by \cite{Amaral2004}, 
\cite{Amaral2007}. It relies on the maximum entropy (ME) method to transfer
information gained at a coarse scale to start inference at a finner scale.

Advances by Dale and Sereno \cite{Dale1993} permitted the inclusion of information 
from structural MRI to construct a Green's function that is specific to the
subject under study. This can be used in a Bayesian approach
as such additional  information can be incorporated in 
the construction of the model and therefore of the likelihood. 
Once a posterior distribution is constructed, it still remains to  
choose an appropriate numerical algorithm in order to extract 
relevant information in the form of expected values or MAP estimates. 
Sato {\it et al} \cite{Sato2004} 
use the Automatic Relevance Determination approach of Neal 
\cite{Neal1997} 
through the Variational Bayes (VB) approximation, (see also 
\cite{bishop2007}) 
while Nummemaa {\it et al.} 
\cite{Nummenmaa2007} analyzed 
essentially the same mathematical 
structure using both VB and Monte Carlo methods. A schematic description of
the VB method we employ is given in section \ref{basis}. 

In sections \ref{maximumentropy} and \ref{priors} we describe the central 
part of this work,  the
method of backward renormalization  priors based on entropic inference. 
We show how to use 
information obtained at a coarse renormalized scale to improve
the starting point on a finner
scale. This can be iterated up from a very coarse description of the system
to the finnest scale.  The presentation is very general and the
particular form of the set of renormalized lattices depends on the application
at hand. 

Validation of the method appears in  section \ref{results} using 
simulated data sets. After calculating the forward model with a set
of known dipole sources, we assess the results comparing the 
performance of the source localization of 2 dipoles 
at random positions and intensities 
using both the original VB method and the multscale approach.
This is done comparing the distance between the localized dipole 
(or first N strongest dipoles) and the \textit{real} one 
(used to generate the simulated data), 
and also the sensitivity of both algorithms (ROC curves) when the noise is increased.
Finally, we applied our method to the problem of source 
localization of a face/mask paradigm in the data available at
http://www.fil.ion.ucl.ac.uk/spm/data/ (SPM - Wellcome Trust Centre for Neuroimaging).
A discussion and conclusions appear in the final section.

\pagebreak

\section{Methods \label{methods}}Since our contribution lies in the proposed method, 
this is the central part of the paper. But first we describe the inverse problem
in section \ref{basis}.
In this part we don't make any advances and follow the work of 
Sato {\it et al.} \cite{Sato2004}). Then, in \ref{maximumentropy}
 we discuss the central part
of our contribution,  the multiscale approach to the prior distribution. 
\subsection {The forward and inverse problems: from cortical sources to  M/EEG data and back
\label{basis}} 
We follow Sato \cite{Sato2004}) using a linear model that has been discussed by \cite{Nunez2006}, \cite{Geselowitz1967} \cite{Mosher1999}, \cite{Mosher1992} 
\cite{Oostendorp1989} (see also \cite{Scharf2010}).
We denote by ${ V}$ the electric potential at 
the scalp surface, by ${J}$ the density of dipole sources that give rise to those potentials
and ${G}$ the Green function that describes the electromagnetic medium. 
It represents the macroscopic 
physiological and
geometrical details of the head known in this area as the \textit{Lead Field}.

Measurements are subject to noise $\xi$ that has been supposed
to be gaussian and independent of space and time, so that for data $V$ and
dipole  sources density $J$ the model is
\begin{equation}
 V =  {G}J + \xi.
\label{linearmodel}
\end{equation}
For $M$ sensors, $T$ the length of the time series of measurements, $N$ the
number of sites of the lattice where the dipoles live, the dimensions
of the matrices $V, J, G$ and $ \xi$ are respectively 
 $M \times T$, $N \times T$, $N \times M$ and $M \times T$. 
Bayes theorem leads to
\begin{equation}
 P(J|VI) = \frac{ P_0(J|I)P(V|JI)}{P(V|I)}. 
\label{posteriorbasic}
\end{equation}
The method that we follow is essentially the variational Bayes
used by Sato {\it et al.} with a twist. In their method, the
 prior distribution 
of the dipole density $P(J_d)=P(\{J_d(r^d(i))\})$ is parametrized by 
a set of parameters $\alpha_d=
\{\alpha_d(i)\}$ where $\Lambda_d$ is the lattice of positions $r^d(i)$
used to represent
the cortex where the dipoles live. $\alpha_d(i)$ is the inverse 
of the variance of the zero mean gaussian also called the precision,
used as prior for the dipole density vector amplitude $J_d(r^d(i))$
at each site $i \in\Lambda_d$. The direction of each  $J_d(r^d(i))$
vector is fixed and normal to the surface representing the cortex at that
point. 
Incomplete knowledge of $\alpha_d$ prompts the use of hyperpriors for
each individual $\alpha_d(i)$. It is reasonable, as they do, to use a set of
Gamma hyperprior distributions for  the set of $\alpha_d(i)$. Integrating
the gaussian priors over the Gamma distributed  $\alpha_d(i)$
leads to t-student distributions. So that at each site, the distributions
of the density belong to the family given by
\bea
P(J|\bar{\alpha}, \gamma)&=&\int_0^\infty \frac{\alpha^{1/2}}{\sqrt{2\pi}} e^{-\frac{\alpha J^2}{2}} \Gamma(\alpha
|\hat{\alpha},\hat{\gamma}) d\alpha\nonumber \\
&=& \sqrt{\frac{\bar{\alpha}}{2\pi \gamma}}\frac{\Gamma(\gamma +\frac{1}{2})}{\Gamma(\gamma)}
\frac{1}{\left[1+\frac{J^2\bar{\alpha}}{2 \gamma}\right]^{\gamma +1/2}}.
\label{tstudent}
\eea
Call collectively the set of hyperparameters of the Gammas $\theta_d^0$.
The inclusion of the data read from the electrodes, 
$V$, by the Variational  Bayes method,
 leads to a dynamics of the hyperparameters
of the gammas, which converges to a final value $\theta_d^f$.
Their method can be succinctly described by the mapping 
\be
\theta_{d}^0=(\hat \alpha_{d}^0,\gamma_{d}^0)
\mathrel{\mathop{\longrightarrow}^{\mathrm{Variational\,\, Bayes}}} 
\theta_d^f=(\hat \alpha_d^f,\gamma_d^f)
\label{renormalphabasic}
\ee
Their starting point is that every $\theta_d^0(r^d(i)) \in  \theta_0^d$ 
is the same, i.e a
 prior of the dipole density spatially invariant over the cortex. 

Our contribution consists of considering a different  choice of the prior 
distributions. For this we use the theory of backward \cite{Caticha2015}
renormalization priors, which are inspired
in the multigrid prior
 approach used by \cite{Amaral2004} to study fMRI. By using 
entropic inference,  we
can use the posterior in a coarse scale to generate
an informed prior in the next
finner scale.
 
\subsection{The Multiscale problem  \label{maximumentropy}}
\begin{figure}[htb]
\centering
\includegraphics[scale=0.8,bb=14 14 578 163]{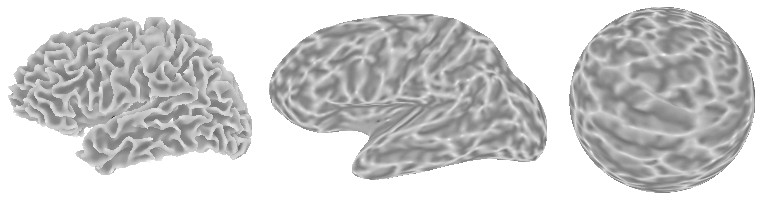}
\caption{Cortex, left hemisphere  (Left) original as obtained from 
a the structural MR image, (center) inflated, (right) spherical. 
Gray scale
code the curvature
in the original image.}
\label{fig:allsurfaces}
\end{figure}

The relevant  variables are the
 dipoles densities $J$ and the 
electric potential (or magnetic fluxes) to be measured at the
electrodes. The space where the variables live
is a representation
of the cortex obtained from structural magnetic resonance data
using Free Surfer image analysis suite, which is documented and freely available for download online 
\newline
(http://surfer.nmr.mgh.harvard.edu/ - \cite{Fischl1999} \cite{Fischl1999a}). 
This representation can be done at different levels of resolution, so that we define a
set of renormalized lattices $\{\Lambda_d\}_{d=0,...D}$. 
Each lattice is composed by sites 
$r^{d}(i)\in \Lambda_d$ with $i=1,....| \Lambda_d|.$
The
particular form of how a coarser or 
renormalized lattice $\Lambda_{d-1}$ is obtained 
from the previous $\Lambda_{d}$ depends on the particular type of problem.
For the problem at hand of EEG dipole sources, the first lattice $\Lambda_0$ was generated by 
introducing an icosahedron in the spherical surface (Figure ~\ref{fig:allsurfaces}) obtained by inflating a brain hemisphere. Deflating the spheres results in a first lattice with 40 faces.
Next, to generate the finer scales lattices $\Lambda_1,..., \Lambda_D$, each trianguar face of the inflated brain was  into 4 triangles. 
More details about the position of dipoles in each face are given in section \ref{simulations}.

At each scale $d$ there is a set of dipole density amplitudes, collectively 
denoted $J_d=\{J_d(r^{d}(i))\}$. 
 We denote the integration measure over the set
of $|\Lambda_d|$ variables by $dJ_d$.

Consider just two consecutive scales, a coarse one $d-1$ and the 
higher resolution $d$. Suppose we have solved the case at the $d-1$ level
and now
we want to solve the problem at scale $d$. This will give a map to be iterated 
from the coarsest to the finest scale. The method to be used follows the simpler case of discrete variables presented in \cite{Caticha2015} is the
maximum entropy (ME), and  the aim is to obtain a distribution 
$P(J_d,J_{d-1},V)$ from a prior distribution $Q(J_d,J_{d-1},V)$. 
The method of choice is the ME because (i) when
constraints are imposed and the prior already satisfies the constraints
ME will result in a posterior equal to the prior and 
(ii) when the results of a measurement
are considered as constraints, ME gives the same results as Bayes
\cite{AdomGArielC2007}. Bayesian usual update can be thought of as a special case of Maximum Entropy
where the constraints arise from the data.

To show (i) we perform a simple Maximum Entropy exercise:
Consider a variable $X$ that takes values $x$ and that $Q(x)$ represents 
our prior state of knowledge. New information is obtained, e.g that the
 expected value $<f(x)>$  is known to have a particular value:
 $<f(x)>=E$. The update from $Q(x)$ to $P(x)$
is done by maximizing 
\be
S[P||Q]=-\int  P(x)\log \frac{P(x)}{Q(x)} dx +\lambda 
\left(\int f(x)dx -E\right)+\lambda_0 \left(\int P(x)dx -1\right)
\ee
The result, after satisfying normalization is the Boltzmann-Gibbs probability 
density $P(x)=Q(x)\frac{\exp \lambda f(x)}{Z(\lambda)}$, with $\lambda$ to be chosen to satisfy  $<f(x)>_P=E$. What is the value of $\lambda$ if 
 $<f(x)>_Q=E$, if the prior already satisfied the constraint? 
It is simple to see that
$\lambda=0$ and $Z(\lambda)=1$, so the reuse of old data in the
from of constraints, again and again
doesn't change the density $Q(x)$  that already satisfies the constraint. While this
sounds trivial, it will be useful since data in Bayes updates can be written
as constraints for Maximum Entropy. 

To prove condition (ii) we follow 
\cite{AdomGArielC2007}. Consider the problem where a  distribution $P(\theta)$
has to be obtained, first, 
from  the knowledge that a measurement of $X$ has yielded
a datum $x'$; second, that prior to the inclusion of such information 
our knowledge of $\theta$ is codified by a distribution $Q(\theta)$
and third, that the relation between $X$ and $\Theta$ is codified by 
a likelihood  $Q(x|\theta)$. Thus we have to consider
$P(x,\theta)$  subject to constraints $\int d\theta P(x,\theta) =P(x)=
\delta (x-x')$. This is not a single constraint, i.e. instead of a single
Lagrange multiplier, we have to consider a function $\lambda(x)$ and maximize 
\bea
S[P||Q]&=&-\int  P(x,\theta)\log \frac{P(x,\theta)}{Q(x,\theta)} dxd\theta
 +\nonumber\\
&&\int \lambda(x) 
\left(\int d\theta P(x,\theta) -\delta(x-x')\right)dx
+\lambda_0 \left(\int P(x)dx -1\right).
\eea
After obtaining the joint density $P(x,\theta)$ by maximizing the entropy,
we can calculate the desired
marginal $P(\theta)=\int dx P(x,\theta)$. The result is
\be 
P(\theta)= Q(\theta|x'),
\ee
where $
Q(\theta|x')=\frac{Q(\theta)Q(x'|\theta)}{Q(x')}$
is the
Bayes  posterior  $Q(\theta|x')$  given by Bayes theorem, which
just follows from the rules of probability. This proves that
maximum entropy as an inference engine justifies the usual Bayes procedure
when the constraint is a datum such as knowing that a measurement of $X$
turned out to give a value $x'$. Maximum entropy is used to 
show that Bayes theorem should be used in the inference process. 

Going back to the EEG problem we consider that the
relevant space is formed by the dipole variables at the
two scales and the electrode potentials.  Thus we seek the maximization of
\bea
S[P||Q]=-\int P(J_d,J_{d-1},V) \log \frac{P(J_d,J_{d-1},V)}{Q(J_d,J_{d-1},V)}
dJ_d dJ_{d-1}dV
\eea
to obtain $P(J_d,J_{d-1},V)$,
which in addition to normalization, is subject to 
\begin{itemize}
\item (A) The marginal $ P(V)=\delta(V-v')$, the measured data is $v'$.
\item (B) $Q(J_{d-1})$ is given, e.g. by 
$\prod_{i\in\Lambda_{d-1}}f(J_{d-1,i}|\theta^{d-1}(r^{d-1}(i)))$ for some parametric family $f(J|\theta)$.
\item (C) Knowledge about the process of renormalization is 
coded by $Q(J_d|J_{d-1})$ (see section \ref{priors}.)
\item Given $J_d$, knowledge of $J_{d-1}$ is irrelevant for $V$: 
$Q(V|J_dJ_{d-1})=Q(V|J_d)$
\end{itemize}
Marginalization and the product rule of probability give
\be
Q(J_d)=\int Q(J_{d-1})Q(J_d|J_{d-1})dJ_{d-1}.
\label{newprior}
\ee
which can be calculated from (B) and (C).
Solving the maximization problem and taking the marginal of the maximum entropy distribution we obtain
\be
P(J_d)=\frac{Q(J_d)Q(v'|J_d)}{Q(v')} \label{posterior}
\ee
is given by what we would have expected, Bayes theorem, 
with the extra important ingredient brought in by equation 
\ref{newprior}, that the prior
at this new scale is obtained by whatever information 
we have on $J_{d-1}$, i.e. $Q(J_{d-1})$ and the renormalization 
procedure $Q(J_d|J_{d-1})$.  

If we wish to restrict the distributions to products of some parametric form , e.g. t-students,  
we have to use  $ Q(J_d|\hat \alpha_d^0,\gamma_d^0)$, from equation
\ref{newprior} given by
\be
 Q(J_d|\hat \alpha_d^0,\gamma_d^0)\approx \int Q(J_{d-1}|\hat \alpha_{d-1}^f,\gamma_{d-1}^f)Q(J_d| J_{d-1}) dJ_{d-1}.
\label{bwr}
\ee
This  is the backward renormalization step.
The idea is to determine
which distribution of $J_d$ in the parametric space of the t-student
distributions is closest to the integral on the right side of  equation 
\ref{bwr}. So a posterior in the coarsest scale $d-1$ induces 
a prior in the $d$ scale. We can start at the lowest resolution 
with the same $(\alpha_0^i,\gamma_0^i)$ initial values at every 
site of the coarsest lattice $\Lambda_0$, obtain via variational 
Bayes the parameters for the posterior of
the $J_0$ from equation \ref{posterior} and proceed
 for the next scales  as represented by the map:

\be(\hat \alpha_{d-1}^f,\gamma_{d-1}^f)\mathrel{\mathop{\longrightarrow}^{\mathrm{BackRenorm}}} (\hat \alpha_d^0,\gamma_d^0)
\label{renormalpha}
\ee
Beginning with a uniform prior at the coarsest scale, that is 
a set of parameters  $(\hat \alpha_0^0,\gamma_0^0)$ uniform over the
lattice $\Lambda_0$, we iterate the mapping 

\be
(\hat \alpha_0^0,\gamma_0^0)\mathrel{\mathop{\rightarrow}^{\mathrm{V B}}}....
\mathrel{\mathop{\rightarrow}^{\mathrm{Back Renorm}}}
(\hat \alpha_d^0,\gamma_d^0)\mathrel{\mathop{\rightarrow}^{\mathrm{V B}}} (\hat \alpha_d^f,\gamma_d^f)\mathrel{\mathop{\longrightarrow}^{\mathrm{Back Renorm}}}
(\hat \alpha_{d+1}^0,\gamma_{d+1}^0)\mathrel{\mathop{\rightarrow}^{\mathrm{V B}}}....
\ee
to finally obtain a posterior at the finest scales described by 
$(\hat \alpha_{D}^f,\gamma_{D}^f)$, the desired
answer to the inference problem. 

Finally it must be emphasized that the marginal 
of the maximum entropy distribution $P(J_{d-1})=\int P(J_d, J_{d-1},V) dJ_ddV$ is
given by Bayes theorem at the coarser level,
$P(J_{d-1})=Q(J_{d-1})Q(V'|J_{d-1}/Q(V')$, showing that the new maximization
of the entropy didn't alter the result obtained 
by the previous step. Reusing the same data within the realm of maximum
entropy is not the same as naively reusing the data using just Bayes theorem
updating. For ME it is harmless, as it imposes a constraint
already satisfied, while for Bayes it represents the belief that
the data were independently re-obtained, leading to a unwarranted
decrease of uncertainty.

\subsection{Backward renormalization priors \label{priors}}
We now investigate the backward renormalization step
given by equation \ref{bwr}. 
Renormalization is seldom a tidy business, and the fact that
the dipoles are vectors and that the direction of their sum is not 
necessarily the
same as the perpendicular to the surface of the cortex, does complicate things
even further. The cortex is represented by triangular faces that are not in the
same plane and thus the renormalized face is not simply related to the finner scale faces. Approximations are needed to advance and suggest a 
specific form for the mapping in display \ref{renormalpha}. 
Numerically we have investigated
this suggestion and found some variations on the theme
 that lead to good results. We can analyze the backward renormalization in 
the following simplified context.
Call
 $\lambda=|\Lambda_{d-1}|/|\Lambda_{d}|$  the ratio 
of degrees of freedom of a lattice at
a stage $d-1$ of renormalization, to the next, finner stage $d$. In this
work $\lambda=1/4.$ 
When going from one 
lattice to a coarser one
 density variables are approximately
renormalized according to 
a scaled block average:
\begin{equation}
J_{d-1}(j)=\lambda\sum_{i(j)}J_{d}(i)
\end{equation}
 where $i(j)$ means that the sum is over the set of 
degrees of freedom at $r^{d}(i)$ 
that are blocked to form the coarser degree of freedom
at position $r^{d-1}(j)$. For independent $J_{d}(i)$  
probability theory leads to the convolution
\begin{eqnarray}
Q_{d-1}(J_{d-1}(j))&=&\int \prod_{i(j)}
\left[dJ_{d}(i) Q(J_{d}(i))\right]
\delta \left( J_{d-1}(j) - \lambda
\sum_{i(j)}J_{d}(i) \right).\nonumber
\end{eqnarray}
Using the  characteristic functions, 
$\Phi(k)= {\cal FT}\left(Q(J)\right)$,
the Fourier transforms of the distributions: 
\begin{equation}
\Phi_{d-1}(k)= \left[ \Phi_{d}(\lambda k)\right]^{1/\lambda}
\end{equation}
So the prior distribution of the dipole at the finner scale position $i$
 can be chosen by inverting a Fourier transform:
\begin{eqnarray}
Q(J_{d}(i))= {\cal IFT}\left( [\Phi_{d-1}( \frac{k}{\lambda}   )]^{\lambda}\right)
\end{eqnarray}
 For distributions stable under additions,
this entails  a simple backward renormalization
of the distribution parameters.
From all this development,  the main information 
we obtained is that the expected value of the 
precision of the gaussian prior of the dipole 
density should  decrease at the new lattice in comparison with that of  
the posterior at the previous coarser lattice.
For the more intricate renormalization we have to consider, a 
further improvement obtained numerically in the
simulations is that not only the variance of the prior should be larger, 
but that at a given lattice the
inferred position of a dipole might be a little off and 
seem to be at a neighboring site. Thus we introduced what we call 
{\it contamination}, by adding variance from the nearest and next nearest sites:
\begin{equation}
\frac{1}{ \bar{\alpha}^{i(j), 0}_{d+1}} = \frac{1}{\bar{\alpha}^{j,f}_d} + 
\frac{1}{3}\sum_{j_{n}} \frac{1}{\bar{\alpha}^{j_n,f}_d}+\frac{1}{9}
\sum_{j_{nn}} \frac{1}{\bar{\alpha}^{j_{nn},f}_d},
\end{equation}
bringing in information from  $j$, the parent site of $i$ as well 
as the nearest neighbors ($j_n$) and the next nearest neighbors ($j_{nn}$) 
of $j$.
This prevents early 
commitment of the position of a dipole. In the average, the precision
scales as the backward renormalization step suggests.
For all the 
sites $i$ in the finner  lattice $\Lambda_d$ that give rise to a given 
renormalization block at $r^{d-1}(j)$, the prior of
 $J(r^d(i))$ will have renormalized parameters inherited from
the block variable distribution of $J(r^{d-1}(j))$.


\begin{figure*}[ht!]
\centering
\includegraphics[scale=0.46,bb=0 0 958 268]{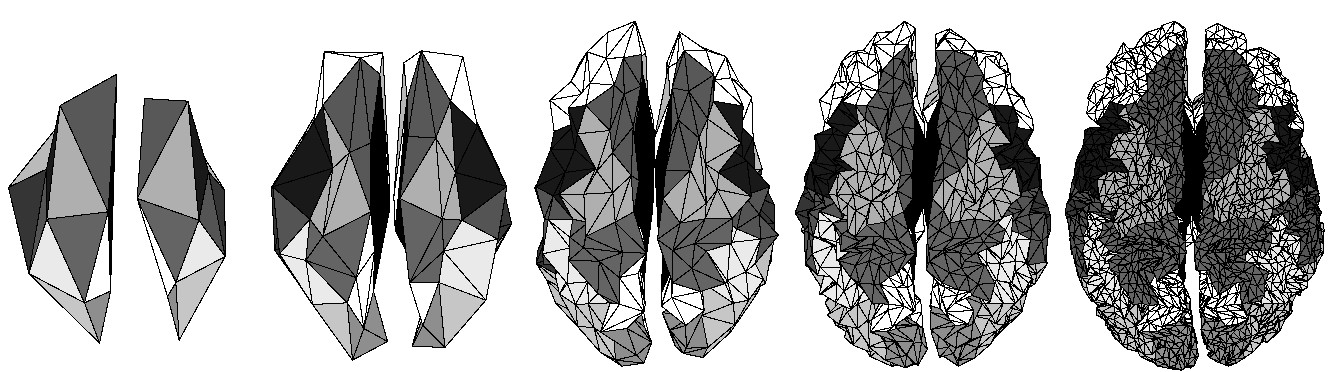}
\caption{Representation of the cortex at different scales. 
Gray scale
is used to represent finer scale balanced surfaces 
derived from the first representation, the icosahedral.
The dipole density vectors are constrained to be at 
a site at the center 
and perpendicular to each triangular face.}
\label{fig:allorders}
\end{figure*}

\pagebreak

\section{Results and discussions
\label{results}}
Depending on the choice of the  initial and final renormalized lattice
several methods can be defined. We denote them by BRVB$_{kk'}$, standing 
for Backward Renormalization Variational Bayes and $k$ indicates the 
initial coarsest scale, $k'$ the finest scale. The final lattice is obtained
by five steps of renormalization. The full method is  BRVB$_{04}$. 
The original Variational Bayes is VB, and since it runs at the finest scale only, 
and is the same as BRVB$_{44}$. We have also compared its performance 
to that of the MNE \cite{Hamalainen1994}.

We first present results obtained by solving the inverse problem for artificial
data generated by the forward problem, obtained  by simulating 
two dipoles in the original lattice generated by Free Surfer. 
Manipulations include (i) the  positions and magnitudes
of the dipoles and (ii) noise corruption of the data.
To quantify the quality of the inference we considered
the distance between {\it real} and localized dipoles (defined in  section \ref{methodvalidation})
estimated by averaging over $\approx 200$ runs.

Since studying the effect of adding noise to cases where BRVB$_{44}$ 
fails is not informative, we considered five different configurations where both 
algorithms performed equally well.
Then, noise level was increased to study the decay
of performance of the different methods.

We end by showing the results of applying the method to publicly available EEG data 
in a face recognition task (http://www.fil.ion.ucl.ac.uk/spm/data/ 
SPM - Wellcome Trust Centre for Neuroimaging).

\subsection{Simulations \label{simulations}}
We start by considering two main active dipoles in the original lattice
 at random positions $r(1)$ and $r(2)$, separated by a distance $L$ ( $ 55 \leq L \leq 65$ mm)
 from each other.
 
Each simulation starts at $t=0$ runs for 
$T = 51$ samples, with each dipole intensity proportional
to  $\sin \frac{t\pi}{T}$. Results shown in the images comes from analysis  
made from data collected at the peak value $t = \, 26$.
  
The forward problem was solved using SPM 
(http://www.fil.ion.ucl.ac.uk/spm/ - Wellcome Trust Centre for Neuroimaging)
and Fieldtrip (http://fieldtrip.fcdonders.nl/ - Centre for Cognitive Neuroimaging of the Donders Institute for Brain, Cognition and Behaviour)
MATLAB toolboxes. They permit calculating the Lead Field using 3-spheres aproximation (for simulations) and 
realistic head model BEM solutions (for analyses of real data).
The Lead Field using 3-spheres was used to generate the potential in the head surface and the signal was then corrupted with 
NSR of $0.1$. We used 128 electrodes positions and MRI template image available with the SPM toolbox.
The inversion methods (MNE, BRVB$_{04}$ through  BRVB$_{44}$ ) were implemented 
in MATLAB. 

As described in section \ref{maximumentropy}, each one of the 5 lattices $\Lambda_0,..., \Lambda_4$ has 40, 160, 640, 2560, 10240 faces, respectively.
The dipoles are positioned using then mean of vertices and faces of the original lattice generated by Free Surfer.
In the spherical surface, it is possible to localize the vertices and faces above each divided face (or the faces of the icosahedron for $\Lambda_0$).
These original faces and vertices have their position in the original folded surface. Each dipole was located in the mean of those vertices 
and oriented as the mean of the normals of those faces. 
This could introduce localization bias since the density of faces in the original lattice is not homogeneous. 
The same set of lattices was generated using balanced representations, but the performance in
all algorithms did not change (results not reported for brevity), so this option was discarded. 
For more information see \cite{Lin2006}.

\begin{figure}[ht]
\centering
\includegraphics[scale=.635,bb=14 14 738 527]{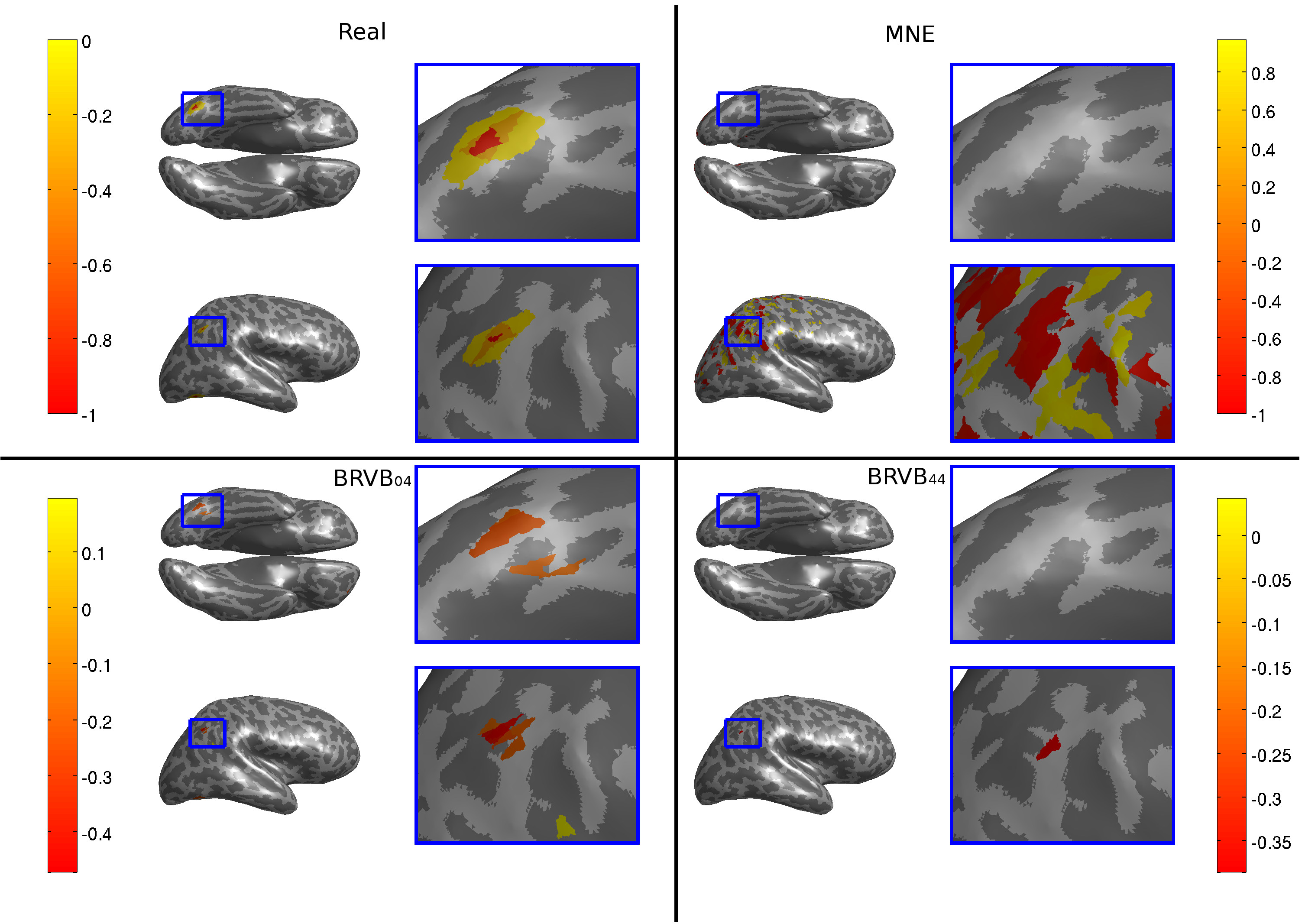}
\caption{
The top left box represents the real current densities
used to simulate the potential, while the others represents the
different localization methods.
Top and bottom line of each box represents the bottom and 
right view of the inflated cortex facing right, respectively.
The first column represents the partially inflated brain 
and the second zooms into  the indicated region in the first column.
Color codes the current intensity, negative facing inwards 
and positive outwards.
The cortex is represented by a lattice  of $\approx 2.8 \times 10^4$ 
triangles and the fifth order grid is  
made up by $10240=2\times 20\times4^4$ possible dipole locations.}
\label{fig:Sources}
\end{figure}
\begin{figure}[htb]
\centering
\includegraphics[scale=1,bb=14 14 460 457]{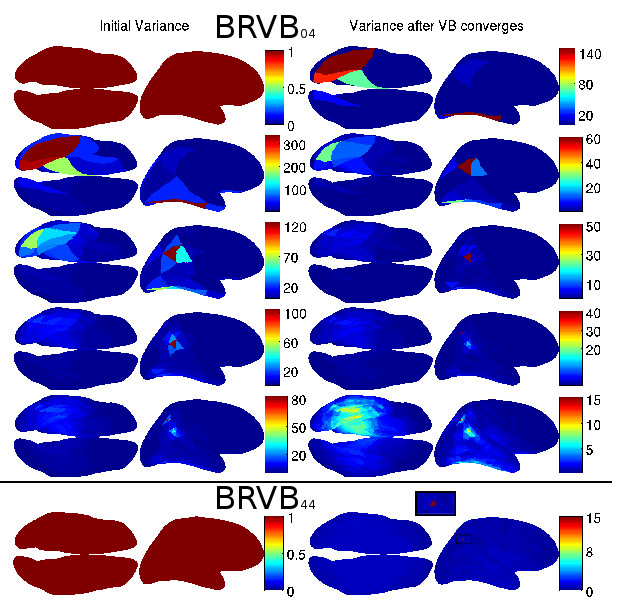} 
\caption{
Evolution of $1/\alpha$ under iteration of the backward renormalization algorithm.
First and third columns shows the bottom view of the cortex surface, while second and fourth columns shows the right hemisphere,
both facing right. First and second columns shows initial variance in each region, while third and fourth
columns shows variance after the VB algorithm converged in that specific grid, with a given initial
value.
First to fifth rows show the stages of BRVB algorithm in each grid 
with: first  BRVB$_{00}$ ($40=2\times 20$ faces), second  BRVB$_{01}$ ($160$)
third  BRVB$_{02}$ ($640$), fourth  BRVB$_{03}$ ($2560$) and fifth 
 BRVB$_{04}$ ($10240$) ($2\times 20 \times 4^d$ faces). 
The last line is the variance in the VB algorithm starting with a hyperparameter $\bar \alpha$
uniform  in the fifth lattice,  BRVB$_{44}$. Notice how aggressive is the convergence in the variance
when the VB method starts with a uniform prior in the last surface, as evidenced by the zoomed square.}
\label{fig:Variance}
\end{figure}

Figure \ref{fig:Sources} shows as an example a particular run. 
In the first line, 
first panel in the left, the {\it real} dipole sources used to generate the
data $V$. 
The following panels
shows the estimated sources using Minimum Norm or MNE,  BRVB$_{04}$ and 
BRVB$_{44}$, respectively.
Figure \ref{fig:Variance} shows the variance or inverse precision
  $1/\alpha_n$ initial value in the two first columns and after convergence
in the two last columns, at each scale for BRVB$_{04}$ in the first 5 lines and in the single grid BRVB$_{44}$ in the last. 

This is a nice example to show because it exemplifies 
the case when the random choice of the location of 
the sources placed one on a deep position, which is known to present a problem since the Lead field is very small. 
This problem has been addressed by \cite{Pascual-Marqui2002} using 
the precision matrix, but here this information is extracted from the data. 
While the BRVB$_{44}$ found only one, the more superficial source, the 
BRVB$_{04}$ was able to find both sources. 
For the two source case, about 10\% of the 100 runs analyzed showed this difference in the behavior for deep sources
and the method was seen to be robust under this condition.

\clearpage
\newpage

\newpage
\clearpage
\newpage

\subsection{Validation of the method \label{methodvalidation}}

The main reason for simulating this problem is that it allows for
comparisons with the {\it real} sources, used to generate the data.
We used mainly the distance between the strongest and second strongest 
localized dipoles.



\begin{figure}[htb]
\centering
\includegraphics[scale=0.85,bb=14 14 506 409]{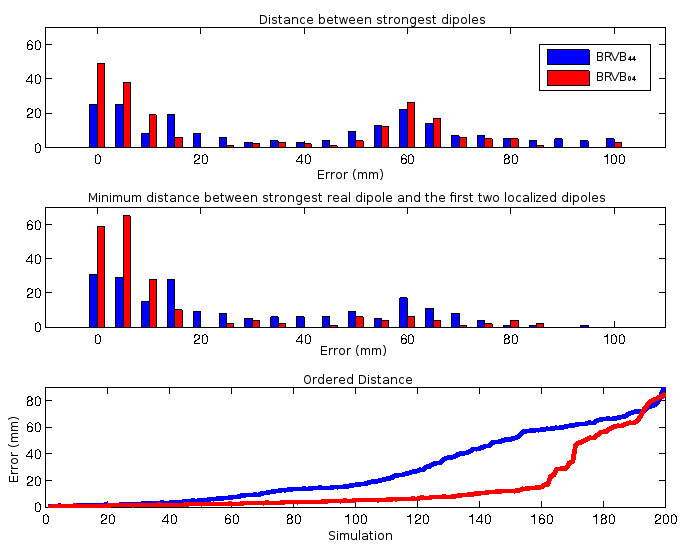} 
\caption{
First line shows the histograms for the distance (mm) between the strongest {\it real} source
 and that found by different algorithms: BRVB$_{04}$ and BRVB$_{44}$.
Second line is the same type of histogram but showing the smallest distance between the strongest {\it real} source
and the two strongest localized dipoles. This way, since we used two sources for generating the field, 
we don't care which one was localized as the strongest, only it's position.
The last line shows the same information as the second histogram, but ordering the errors in ascendant fashion.
Obtained from inference made on 200 different forward problems, with random positions and
distances between the two dipoles randomly chosen between 55 and 65 mm.}
\label{fig:SimSuperficies}
\end{figure}

To begin we compare the distances in mm between {\it real} and localized dipoles in the 200 simulations.
As we can see in Figure \ref{fig:SimSuperficies}, we first compare the strongest dipole found by the methods BRVB$_{44}$ and BRVB$_{04}$ and the strongest {\it real} dipole.
We also compare it to the minimum between the distances of the first and second strongest localized ones,
ignoring differences in amplitudes. It is interesting to see that even in this case there is a peak at 60 mm, specially for the 
BRVB$_{44}$. 

Finally we show the ordered errors for all simulations, from BRVB$_{44}$ to BRVB$_{04}$, evidencing the improvement in the 
use of each additional localization.


\newpage

We also analyzed how different algorithms fare under the addition of different levels of noise.  





We identified 5 simulations where both the BRVB$_{44}$, 
BRVB$_{34}$ and BRVB$_{04}$ obtained similar good results 
from data corrupted at low noise-to-signal ratios (NSR=0.1). In order to compare
runs with similar good results, we have to restrict to cases where the sources are located in more
superficial regions, since the performance for sources in deep locations are quite different with the
single scale method not even identifying a result to be compared. We added
uncorrelated gaussian noise of zero mean and variance $\sigma^2$
to each of the $M$ components (electrodes) of the vector $V$ of data voltages so that 
\begin{equation}
 NSR = \frac{M \: \sigma^2}{V'V}.
\end{equation}

\begin{figure}[ht!]
\centering
\includegraphics[scale=0.85,bb=14 14 474 323]{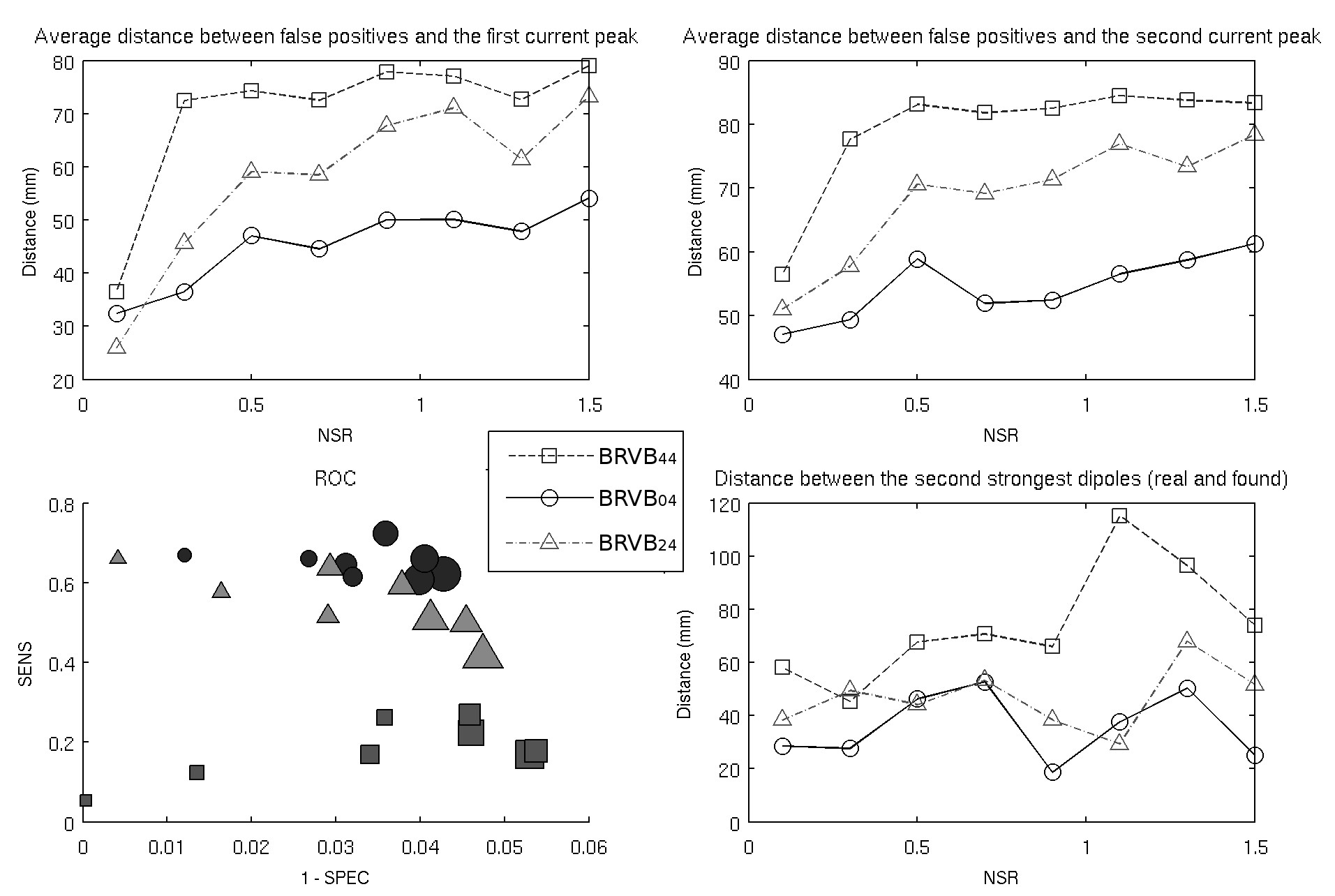} 
\caption{Performance of the different algorithms. The horizontal axis represents the noise to signal ratio
(apart from the ROC space). First line : average distance between false positives and the most intense
real dipole on the left, and the second most intense on the right. 
Second line : ROC space on the left (the size of the markers represents the intensity of noise),
and average distance between the second most intense true positive and the second most intense real dipole on the right.}
\label{fig:SimNSR}
\end{figure}

Finally, in Figure \ref{fig:SimNSR}, besides the error distance in localization, we also plotted the ROC space for the three algorithms,
and the curves for the distance between the average false positives positions to the {\it real} dipoles.
Notice how the the $BRVB_{04}$ has a clear superior resistance to the increased noise.


\newpage

\subsection{Real Data from EEG}

\begin{figure}[ht!]
\centering
\includegraphics[scale=.5,bb=14 14 863 364]{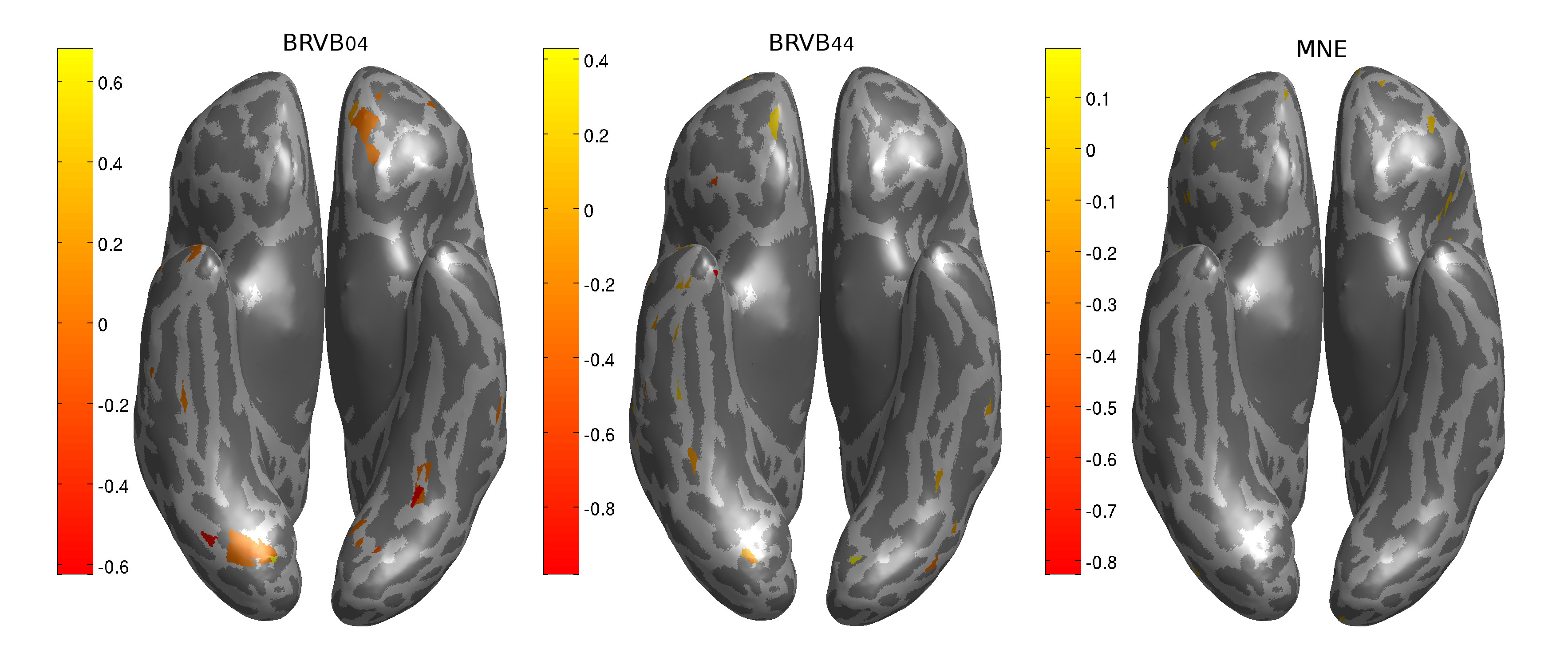}
\caption{Bottom view of the cortex looking up (most significant locations are on the bottom in that instant). 
From left to right : BRVB$_{04}$, BRVB$_{44}$ and MNE at approximately 170 ms. }
\label{fig:faces}
\end{figure}

We applied the BRVB$_{04}$ and the BRVB$_{44}$ in real data available at 
\newline
http://www.fil.ion.ucl.ac.uk/spm/data/mmfaces/
(SPM - Wellcome Trust Centre for Neuroimaging). The experiment consists of 128 electrodes set in 
a Face / No-Face stimulus. We preprocessed the data as specified in the tutorial for source reconstrucion in SPM, and 
computed the difference between the average of each condition. The resulting
scalp potential was used for the source localization. The sources shown in Figure \ref{fig:faces} were found by BRVB$_{04}$,
BRVB$_{44}$ and MNE. Although they are in general consistent with the literature for this experiment \cite{Halgren2000}, 
the location of the sources as well the intensity is clearly different. Further studies are necessary to understand the nature of this difference
in this specific protocol.


\section{Conclusion}



The main contribution of this paper is  to use  a Maximum Entropy
inferential chain of projections  to systematically
transfer posterior information from one coarse scale to the
prior of a finner scale. 
This approach is not restricted to M/EEG or fMRI imaging methods, but 
should apply to inference about the localization of sources
 in any spatially extended system and thus has a
potentially wide scope of applications.
We analyzed the behavior of the hierarchical Bayesian approach for solving the 
M/EEG inverse problem. This was introduced into this context by \cite{Sato2004} who  used one coarser
grid to restrict the search in a finer scale, without propagating
the converged variance to the next scale. As shown by Nummenmaa {\it et al}
\cite{Nummenmaa2007}, the 
Variational Bayes approach is very sensitive to the initial values of the hyperparameters.

Our  multiscale approach was inspired by the fMRI work of \cite{Amaral2004} to systematically 
construct prior distributions. An important difference is that the M/EEG 
data introduces a further complication by not being a scalar as in  the fMRI case, leading us to 
defining {\it a priori} directions for the dipoles, or risk having to deal with a nonlinear problem 
with a too large number of degrees of freedom.  This simplicity permitted the more sophisticated 
formulation based on the backward renormalization group for discrete variables done in \cite{Caticha2015}
Our method permits setting 
the initial value for the hyperparameters in the VB approach on a finer scale from the value obtained 
after VB convergence in the coarser previous scale. The method being robust to choices in the coarsest scale. However the renormalization group anlysis of this harder problems remains incomplete and further
work in this direction will follow.

Second, we have done extensive simulations to validate the method. 
We presented results in simulated data that suggests that this approach is a valid aid
to increase the precision of the localized dipoles and also to increase the performance in the presence
of noise. The backward renormalization priors permitted  a more systematic localization of  deep
sources far from the skull.

Third, as mentioned by Nummenmaa {\it et al} \cite{Nummenmaa2007} in the discussion, there appears to be a natural trade-off between choosing a 
method providing smoother but unique solution, and the hierarchical approach with better spatial 
resolution and a multitude of candidate solutions. We claim that our method might represent a good
direction in finding a compromise between both solutions.

We stress that the main advantages of a Maximum Entropy or Bayesian approach is the 
clear identification of the often hidden assumptions underlying an algorithmic approach. 
This permits concentrating on the different pieces needed to solve the puzzle.
It illustrates the fact that prior information not only goes into the prior but can improve the likelihood. 
While we need a better electromagnetic model of the brain as well as
understanding the noise processes, here 
we just looked at the prior, but this can certainly be improved. A natural idea is to improve our method with  
anatomical prior information as used by \cite{Ou2010}.

{\bf   Acknowledgements} We thank Ariel Caticha, Selene Amaral and Said 
Rabbani for discussions on different aspects of inference
and neuroimaging. This work received support from CNAIPS-USP and CNPq.

    \bibliography{barcat}
\end{document}